# A Passive Daytime Colored Radiative Cooler with DBR-Engineered Color Selectivity

Rajib Lochan Ghadei, Rohit Gupta and Rishi Maiti*
Department of Physics, Indian Institute of Technology Guwahati, Assam, 781039, India
*Corresponding author's email: rmaiti@iitg.ac.in

## Abstract:

Passive daytime colored radiative cooling (PDCRC) offers a promising route for achieving efficient thermal management while providing attractive structural colors. However, simultaneously maintaining strong daytime cooling and vivid coloration remains a challenge. Here in, we propose a lithography-free PDCRC comprising a polydimethylsiloxane (PDMS) emitter, $SiC/SiO_2$ distributed Bragg reflector (DBR), $MgF_2$ spacer, and Ag back reflector. The DBR and $MgF_2$ spacer are designed to provide selective spectral absorption in the visible range for cyan, magenta, and yellow (CMY) color generation, while the PDMS layer enables strong thermal emission within the atmospheric transparency window (8–14 µm). The proposed structure achieves an average emissivity of 92% in the atmospheric window and an average reflectivity of 93.3% over 0.75–6 µm, as calculated using the finite-difference time-domain (FDTD) method and validated by the transfer matrix method (TMM). CIE 1931 chromaticity analysis confirms high-contrast CMY colors with low color differences from the corresponding standard subtractive primary colors. Under practical daytime conditions with a heat-transfer coefficient of 6 W $m^{-2}$ $K^{-1}$, the proposed cooler delivers cooling powers of approximately 140–145 W $m^{-2}$ and a temperature reduction of about 12°C below ambient temperature. The color characteristics remain stable over a broad range of incident angles, with only a gradual reduction in color intensity at larger angles, while the spectral response remains nearly insensitive to TE and TM polarizations. Furthermore, simulations based on meteorological conditions of different global cities demonstrate the robust cooling capability of the proposed structure under diverse environmental conditions. The combination of high cooling performance, vivid coloration, simple multilayer architecture, and fabrication feasibility makes the proposed PDCRC promising for energy-efficient thermal management in buildings, vehicle coatings, wearable devices, and other outdoor applications.

## 1. Introduction

Global warming and the increasing frequency of extreme heat events have substantially increased the demand for space cooling. The growing adoption and use of air-conditioning (AC) systems, particularly in regions such as Europe and countries such as the United States, India, and China, have placed an increasing burden on electricity consumption and peak power demand. As cooling requirements continue to increase with increasing ambient temperatures, conventional cooling technologies alone may not provide a sustainable and energy-efficient solution for meeting future cooling needs. Therefore, the development of alternative cooling technologies that can reduce electricity consumption while providing effective thermal management is becoming increasingly important for a sustainable future.

Passive daytime radiative cooling (PDRC) has emerged as a promising cooling technology that operates without external energy input, unlike air conditioners and refrigerators. Rather than replacing existing cooling systems, the PDRC offers an environmentally friendly and energy-efficient approach to supplement cooling requirements during periods of intense heat. Under direct sunlight, passive radiative coolers have been demonstrated to reduce surface temperatures by approximately 7–15°C below ambient conditions without consuming electricity [1], [2]. The cooling effect originates from a combination of solar reflection and thermal radiation emission from the cooling surface. During nighttime, cooling occurs through the emission of thermal radiation from the surface directly into the extremely cold outer space (~3 K). This process is particularly effective within the atmospheric transparency window (8–14 μm), where the atmosphere exhibits high transmittance and minimal absorption of thermal radiation [3]. As a result, thermal energy emitted by the surface can escape into outer space with negligible atmospheric attenuation, enabling passive cooling. However, nighttime emission alone is insufficient for achieving efficient cooling under daytime conditions. During the day, incident solar radiation within the solar spectrum (0.3–2.5 μm) can be strongly absorbed by materials, leading to significant heat accumulation. Therefore, effective daytime radiative cooling requires simultaneous suppression of solar heating through high reflectivity within the solar spectrum while maintaining strong thermal emission within the atmospheric transparency window [4]. By minimizing solar absorption and maximizing thermal radiation loss, net cooling can be achieved even under direct sunlight. In addition to cooling performance, aesthetic appearance has become an important consideration for applications such as buildings, vehicles, consumer products, and decorative coatings [5], [6]. Conventional radiative coolers typically exhibit white, silver, or metallic appearances owing to their broadband solar reflectance, which often limits their visual appeal and architectural integration [7], [8]. To address this challenge, passive daytime colored radiative cooling (PDCRC) has emerged as an attractive approach that combines efficient radiative cooling with structural coloration. This is achieved through selective absorption within the visible wavelength range (0.4–0.7 μm), while simultaneously preserving high solar reflectance and strong thermal emissivity. Consequently, the design of a PDCRC requires precise engineering of electromagnetic waves across three distinct spectral regions: the visible spectrum for color generation, the solar spectrum for minimizing heat absorption, and the atmospheric transparency window for maximizing thermal emission. Achieving an optimal balance among these competing requirements remains one of the key challenges in the development of high-performance colored radiative cooling systems [9], [10].

The development of radiative cooling technologies accelerated following the first experimental demonstration of passive daytime radiative cooling by Raman et al. in 2014, where a temperature reduction of approximately 5°C below ambient temperature was achieved under direct sunlight [11]. Since then, extensive efforts have been devoted to the development of high-performance radiative coolers using hybrid micro/nanoporous polymers [12], multilayer thin films [13], bio-inspired material designs [14] and metasurfaces [15] to achieve efficient sub-ambient cooling. Subsequently, increasing attention has been directed toward passive daytime colored radiative cooling (PDCRC), which combines thermal management with aesthetic appearance. In 2013, Zhu et al. theoretically proposed the first color-preserving radiative cooler based on a periodic quartz array, demonstrating a daytime temperature reduction of up to 14.4°C [16]. Later, Lee et al. experimentally realized a thin-film optical resonator-based colored radiative cooler in 2018, achieving sub-ambient cooling of 3.9°C [17]. In 2019, Sheng et al. proposed an optical Tamm resonance-based colored radiative cooler that theoretically exhibited cooling powers of 44–52 W m$^{-2}$ for the three subtractive primary colors. Their design achieved a temperature reduction of 5–6°C below ambient conditions at a non-radiative heat transfer coefficient ($h_c$) of 6 W m$^{-2}$ K$^{-1}$ [18]. In 2022, Yu et al. reported a microparticle-polymer-based colored radiative cooler capable of experimentally achieving a sub-ambient cooling of 4 K under a solar irradiance of 1000 W m$^{-2}$ [19]. Subsequently, Pirouzfam et al. developed a plasmonic-assisted $SiO_2$–$TiO_2$ periodic structure in 2023, obtaining a cooling power of approximately 60 W m$^{-2}$ [20]. In 2025, Xie et al. designed a hybrid photonic radiative cooler consisting of a $SiO_2$ hole-array emitter integrated with a metal–insulator–metal reflector, demonstrating a theoretical cooling power of up to 139 W m$^{-2}$ at an ambient temperature of 300 K [21]. Despite these advances, achieving high cooling performance together with vivid color appearance remains a significant challenge. In most colored radiative coolers, the introduction of visible coloration often compromises cooling efficiency because color generation generally requires additional optical structures that influence solar reflectance and thermal emissivity. Multilayer coatings, resonant cavities, and metasurfaces can provide attractive coloration; however, they frequently involve complex fabrication processes and increased manufacturing costs. Therefore, maintaining an optimal balance among color purity, solar reflectance, thermal emissivity, and fabrication simplicity remains an important research objective.

Motivated by these challenges, we propose a lithography-free passive daytime colored radiative cooler consisting of a single thick emissive layer combined with a compact color-selective reflector. The numerical simulation is performed using the FDTD method followed by the TMM method. The proposed structure employs polydimethylsiloxane (PDMS), as the top emissive layer, followed by SiC/$SiO_2$ distributed Bragg reflector (DBR) layers and an $MgF_2$ spacer layer for color generation and spectral tuning. A silver (Ag) film is incorporated at the bottom as a highly reflective back reflector. Owing to its high transparency in the visible region, the PDMS layer does not interfere with color formation while simultaneously serving as an efficient thermal emitter due to its strong emissivity within the atmospheric transparency window (8–14 μm). The proposed design achieves an average emissivity of approximately ~92% across the atmospheric transparency window and an average reflectivity of ~93.3% within the near-infrared region (0.75–6 μm). By carefully engineering the multilayer structure, narrow spectral absorption bands are introduced in the visible region to generate high-contrast cyan, magenta, and yellow (CMY) colors without significantly affecting the cooling performance. As a result, cooling powers of ~145 W $m^{-2}$ are achieved for all three colors, together with a temperature reduction of approximately 12°C below an ambient temperature of 30°C. The systematic studies of the angular and polarization dependent optical response reveal that variations in incident angle primarily affect color contrast and cooling performance while preserving the original color identity. The cooling characteristics are also evaluated under realistic environmental conditions using meteorological data obtained from the NASA Earth Climate database for different global locations. The obtained results demonstrate the robustness, environmental adaptability, and practical applicability of the proposed PDCRC design, highlighting its potential for energy-efficient thermal management in buildings, vehicle coatings, wearable technologies, and other outdoor electronic applications.

## 2. Multilayer Design and Numerical Methods

### 2.1. Design and Simulation of the Colored Radiative Cooler

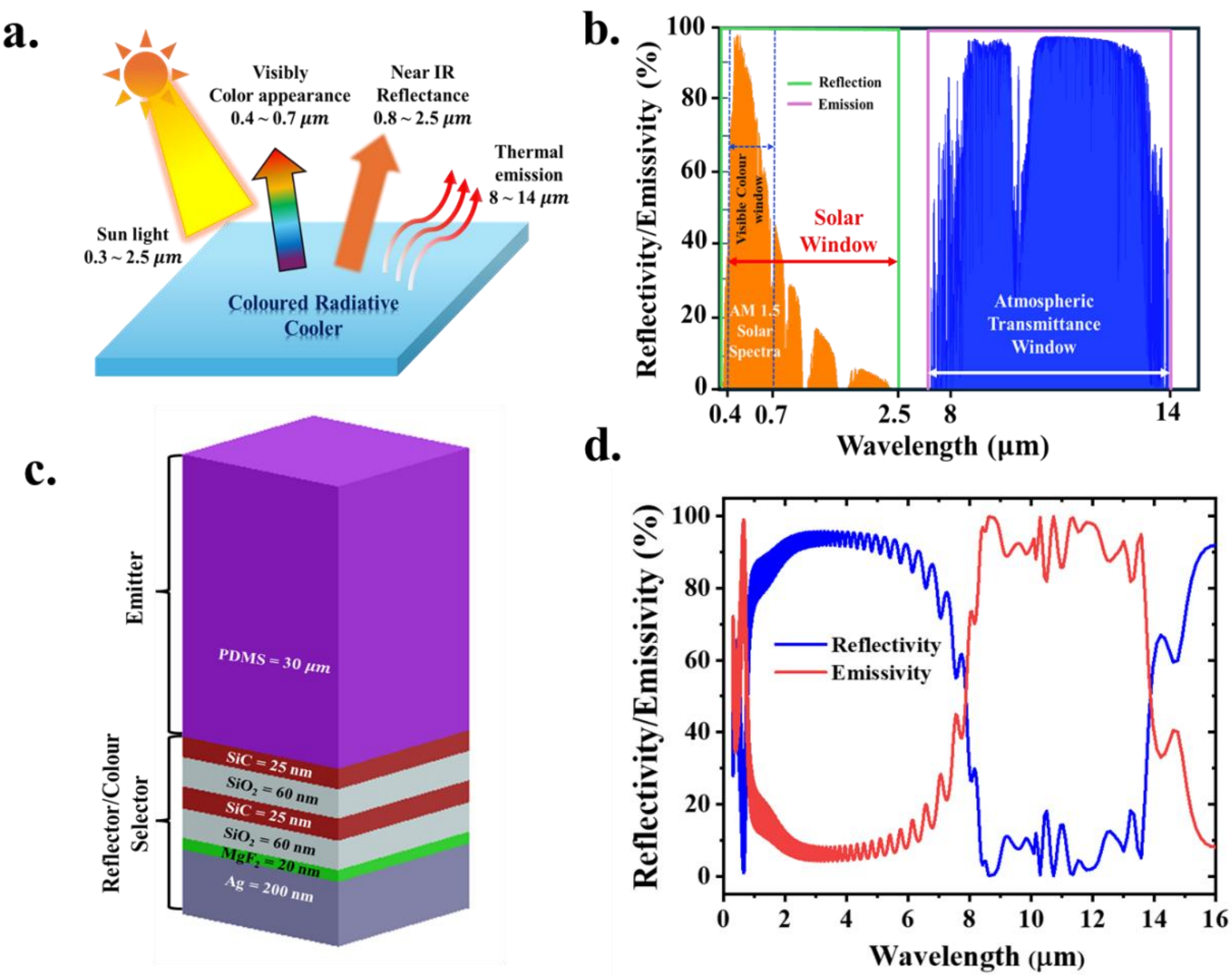

**Fig. 1.** (a) Schematic of passive daytime colored radiative cooler (PDCRC). (b) Spectral ranges highlighting the visible band (0.4–0.7 μm), solar window (0.3–2.5 μm), and atmospheric transparency window (8–14 μm). (c) Three-dimensional schematic of the PDCRC, consisting of a top emitter layer and underlying reflector/color-selector layers. (d) Reflectivity and emissivity spectra of the PDCRC, showing high reflectivity across the solar spectrum with a reflection dip in the visible range and high emissivity (~92%) within the atmospheric window (8–14 μm).

Fig. 1(a) illustrates the operating principle of colored radiative cooling. Since most of the solar spectrum lies within 0.3–2.5 μm, the structure is designed to reflect this range while enabling selective absorption in the visible band (0.4–0.7 μm) for color generation. Simultaneously, thermal radiation is emitted within the mid-infrared region (8–14 μm), enabling passive heat dissipation to outer space through the atmospheric transparency window. Fig. 1(b) shows the ideal spectral response, characterized by high solar reflectivity and strong emissivity within the atmospheric window. Visible-band spectral selectivity is required for color generation.

A lithography-free design of the passive daytime colored radiative cooler (PDCRC) is presented in Fig. 1(c). The structure spans 5 μm × 5 μm and consists of a 30 μm thick polydimethylsiloxane (PDMS) top layer acting as the thermal emitter. To achieve simultaneous solar reflection and visible selectivity, a multilayer stack is incorporated beneath the PDMS. This includes a distributed Bragg reflector (DBR) formed by two pairs of SiC (25 nm) and $SiO_2$ (60 nm), followed by a 20 nm $MgF_2$ spacer and a 200 nm Ag back reflector. The structure is supported on a silicon substrate, which is omitted from the simulation as it does not affect the optical response. The optical response is evaluated using the finite-difference time-domain (FDTD) method over 0.3–16 μm with a broadband plane-wave source. A non-uniform mesh with a minimum mesh size of 2.5 nm is applied along the x, y, and z directions. The simulation is performed for 2000 fs with a time step of 0.0047 fs and a stability factor of 0.99. A mesh accuracy level of 4 is used to ensure convergence. Perfectly matched layer (PML) boundary conditions are applied along the z-direction, while periodic boundary conditions (PBCs) are imposed along x and y [22]. Transmission is negligible due to the optically thick Ag layer, and the incident radiation is partially reflected and absorbed. Reflectivity (R) and transmissivity (T) are obtained using frequency-domain power monitors. The absorptivity is calculated as [23]:

$$A(\%) = (1 - R - T) \times 100 \qquad (1)$$

According to Kirchhoff's law of thermal radiation, emissivity equals absorptivity at thermal equilibrium; therefore, the emissivity of the PDCRC is taken equal to its absorptivity. Refractive index data of the above materials are obtained from the literature [24], [25], [26], [27], [28]. The SiC, $SiO_2$, and $MgF_2$ layers can be deposited by RF magnetron sputtering, followed by thermal evaporation of the Ag layer and spin-coating of PDMS. Fig. 1(d) shows that the proposed structure closely approaches the ideal spectral response. High average reflectivity is maintained across the solar spectrum, minimizing solar absorption, while strong emissivity (~92%) is achieved within the atmospheric window. A distinct reflection dip in the visible range induces wavelength-selective absorption, which determines the perceived color.

## 2.2. Validation of FDTD Results Using the Transfer Matrix Method (TMM)

The FDTD simulation results are validated using the transfer matrix method (TMM). Fig. 2(a) illustrates the TMM framework applied to the multilayer structure, where each layer is assumed to be homogeneous, isotropic and tangential components of electric and magnetic fields are continuous at interfaces [29]. The surrounding media above and below the stack are assumed to be air with a refractive index of $n_0 = 1$. A normally incident plane wave is considered for the analysis. The wavelength-dependent complex refractive indices of PDMS, SiC, $SiO_2$, $MgF_2$, and Ag are taken from literature [24], [25], [26], [27], [28], and the corresponding layer thicknesses ($t_1$–$t_7$) are defined as shown in Fig. 2(a). The TMM evaluates the optical response of the multilayer system by accounting for multiple reflections and interference effects arising at each interface. Using the combined effects of layer thickness, refractive index, wavelength, and incidence conditions, the spectral absorptivity/emissivity of the entire structure is calculated over the wavelength range of 0.3–14 μm. In the TMM formulation, the propagation of the electromagnetic wave through each layer is described using a characteristic matrix. For the $j$-th layer, the matrix is expressed as [30]:

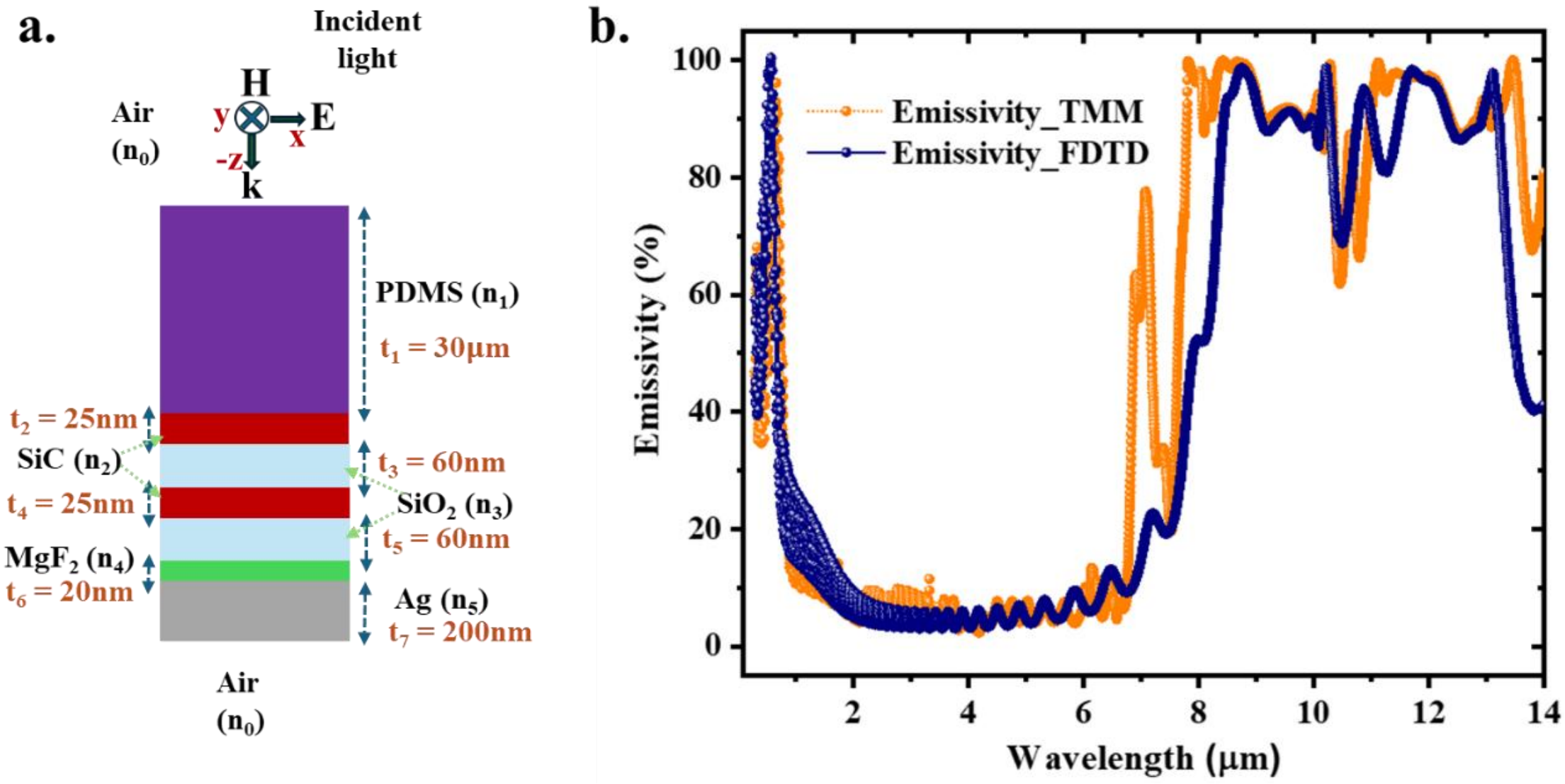

**Fig. 2.** (a) Schematic of the transfer matrix method (TMM) used to compute the net emissivity from the multilayer PDCRC based on individual layer thicknesses and refractive indices. (b) Comparison of emissivity spectra obtained from TMM and finite-difference time-domain (FDTD) simulations. The two approaches show close agreement across the solar window (0.3–2.5 μm) and the atmospheric transparency window (8–14 μm).

$$M_j = \begin{bmatrix} \cos(\delta_j) & \frac{i}{q_j}\sin(\delta_j) \\ iq_j\sin(\delta_j) & \cos(\delta_j) \end{bmatrix} \tag{2}$$

where $\delta_j = \frac{2\pi}{\lambda} n_j t_j$ represents the phase thickness, with $n_j$ and $t_j$ being the complex refractive index and thickness of the $j$-th layer, respectively. For normal incidence, the optical admittance is given by $q_j = n_j$.

The overall transfer matrix of the multilayer system is obtained by multiplying the individual layer matrices:

$$M = \prod_{j=1}^{N} M_j \tag{3}$$

From the total transfer matrix, the reflection and transmission coefficients are determined as:

$$r = \frac{(M_{11} + M_{12}q_s)q_0 - (M_{21} + M_{22}q_s)}{(M_{11} + M_{12}q_s)q_0 + (M_{21} + M_{22}q_s)} \tag{4}$$

$$t = \frac{2q_0}{(M_{11} + M_{12}q_s)q_0 + (M_{21} + M_{22}q_s)} \tag{5}$$

where $q_0$ and $q_s$ denote the optical admittances of the incident and substrate media, respectively. The reflectivity and transmissivity are then obtained as:

$$R = \mid r \mid^2, T = \frac{\mathrm{Re}(q_s)}{\mathrm{Re}(q_0)} \mid t \mid^2 \tag{6}$$

The absorptivity (and hence emissivity) is calculated using:

$$A(\lambda) = 1 - R(\lambda) - T(\lambda) \tag{7}$$

Using this formulation, the spectral absorptivity/emissivity of the multilayer structure is evaluated over the wavelength range of 0.3–14 μm. This analytical approach provides a reliable benchmark for validating the FDTD simulation results.

Fig. 2(b) shows a comparison between the emissivity spectra obtained from TMM and FDTD. Excellent agreement is observed across the UV–visible–near-infrared (NIR) region (0.3–6 μm), confirming the accuracy of the FDTD model. In the atmospheric transparency window (8–14 μm), both approaches predict high emissivity, reaching approximately 93.2% and 92% for TMM and FDTD, respectively. Minor deviations from 6.5–8 μm can be attributed to differences in numerical discretization, mesh resolution and finite-time convergence effects in the FDTD simulations [31]. The transfer matrix method (TMM) provides an efficient and exact solution for ideal one-dimensional multilayer structures. However, the finite-difference time-domain (FDTD) method is employed to perform full-wave simulations, enabling access to field distributions and ensuring applicability to more general geometries beyond ideal planar assumptions. The agreement between TMM and FDTD validates the accuracy of the numerical model.

## 3. Mechanisms of Spectral Control for Color Generation and Radiative Cooling

### 3.1. Material-Dependent Optical Properties and Spectral Response

Figures 3(a–d) present the wavelength-dependent complex refractive indices of SiC, $SiO_2$, Ag, and PDMS, which govern the spectral response of the multilayer structure. In the NIR region (≈ 0.75–6 μm), SiC, $SiO_2$, PDMS, and $MgF_2$ exhibit negligible extinction coefficients (k ≈ 0), indicating minimal intrinsic absorption and allowing efficient transmission of incident light through these layers. In contrast, Ag exhibits a large extinction coefficient over this range, resulting in strong reflection at the metal interface with only minor absorption losses. This combination leads to high reflectivity in the NIR region, reaching ~93.3% over 0.75–6 μm. In the mid-infrared region (8–14 μm), all constituent materials exhibit increased extinction coefficients, leading to enhanced absorption within the multilayer stack. This collective absorption contributes to a high emissivity of 92% within

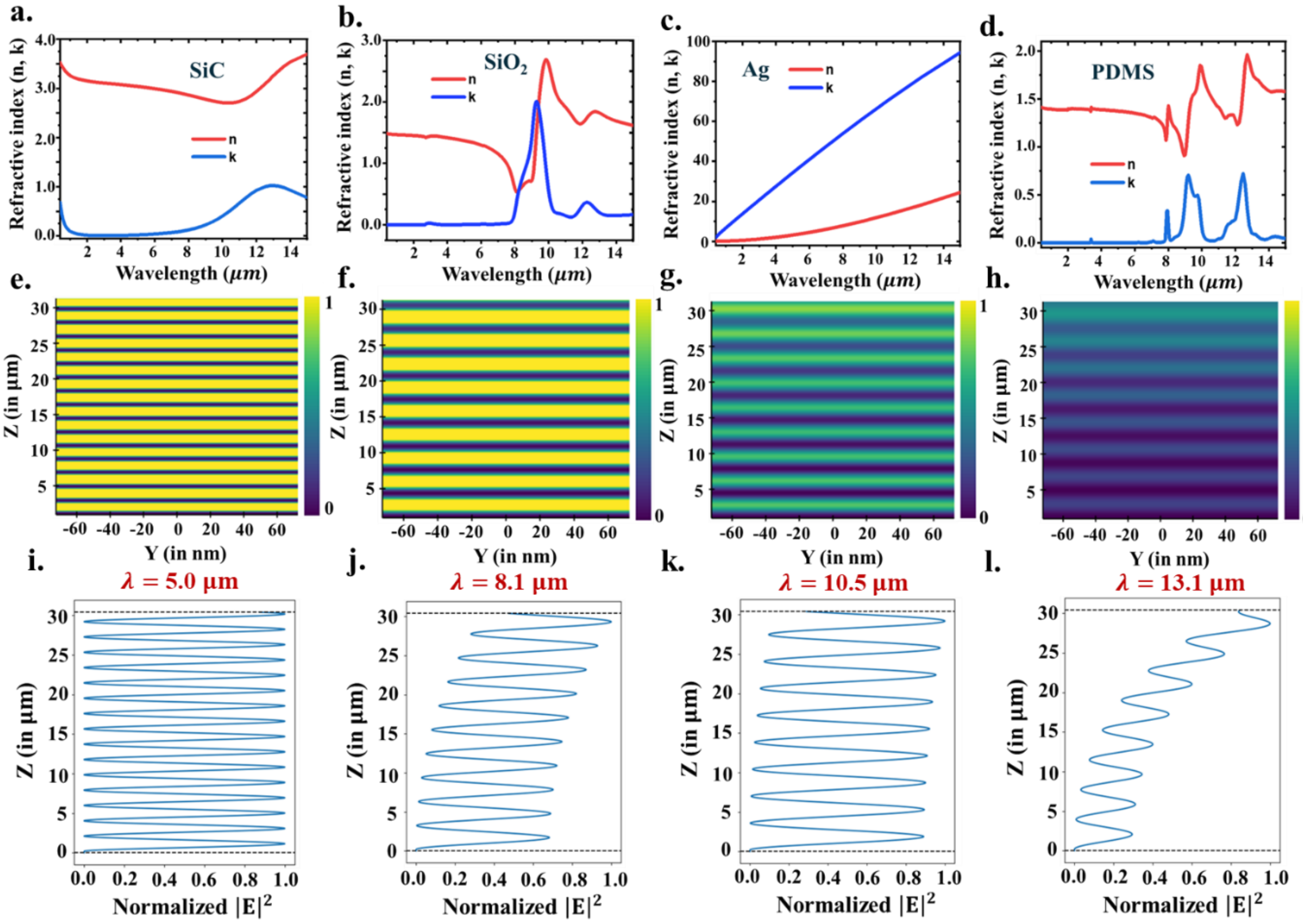

**Fig. 3.** (a–d) Wavelength-dependent complex refractive index (n, k) of the constituent materials taken from literature: (a) SiC, (b) $SiO_2$, (c) Ag, and (d) PDMS, illustrating their optical properties across the solar and mid-infrared spectral ranges. (e–h) Electric field intensity ($|E|^2$) distributions obtained from FDTD simulations at wavelengths of (e) 5.0 μm, (f) 8.1 μm, (g) 10.5 μm, and (h) 13.1 μm. (i–l) Corresponding electric field intensity ($|E|^2$) distributions calculated using the transfer matrix method (TMM) at (i) 5.0 μm, (j) 8.1 μm, (k) 10.5 μm, and (l) 13.1 μm, showing good agreement with FDTD results.

the atmospheric transparency window, which is essential for effective radiative cooling. In the visible range, selective absorption arises primarily from the finite extinction coefficient of SiC in the 0.4–0.75 μm range. This absorption, combined with interference effects governed by the DBR stack and the $MgF_2$ spacer thickness, enables wavelength-selective reflectance, thereby producing the observed structural color.

## 3.2. Electric Field Distribution and Wavelength-Dependent Energy Localization

The spectral behaviour across the wavelength range of 0.3–14 μm is further analyzed through the spatial distribution of the electric field intensity ($|E|^2$) obtained from finite-difference time-domain (FDTD) simulations and cross-validated using the TMM. In the TMM framework, the electric field within each layer is expressed as a superposition of forward- and backward-propagating waves along the propagation (z) direction [32], [33]:

$$E_j(z) = A_j e^{ik_j z} + B_j e^{-ik_j z} \quad (8)$$

where $A_j$ and $B_j$ are the complex amplitudes of the forward and backward waves, respectively, and $k_j = \frac{2\pi}{\lambda} n_j$ is the complex wavevector in the $j$-th layer. The electric field intensity along the depth of the multilayer structure is given by:

$$| E_j(z) |^2 = E_j(z)\, E_j^*(z) \quad (9)$$

The wavelength-dependent field distributions provide insight into the transition from reflective to emissive behaviour in the multilayer structure. At λ = 5.0 μm (Fig. 3(e,i)), corresponding to the high-reflectivity regime, the field exhibits strongly confined and closely spaced oscillations arising from multiple internal reflections between the DBR and the metallic back reflector. The low absorption in this region results in minimal energy dissipation, leading to pronounced standing-wave patterns. As the wavelength increases to λ = 8.1 μm (Fig. 3(f,j)), the oscillation spacing becomes less dense, indicating reduced effective reflection and the onset of increased absorption within the structure, consistent with interference-driven field redistribution in multilayer systems [30], [32]. At λ = 10.5 μm and 13.1 μm (Fig. 3(g,k) and (h,l), respectively), the field intensity progressively decays along the propagation direction, demonstrating strong attenuation due to material absorption [33]. The diminished oscillatory behaviour and monotonic field decay confirm efficient electromagnetic energy dissipation, which directly contributes to high emissivity in this spectral range. The consistently similar trends observed in both TMM and FDTD results support this interpretation. The field is strongly confined in the reflective region due to interference, whereas it decays in the emissive region due to absorption, thereby enhancing emissivity and effective radiative cooling.

# 4. Results and Discussion

## 4.1. Structural Color Design and Tuning via DBR and Spacer Thickness

Radiative cooling structures are primarily designed to achieve high solar reflectivity in the 0.3–2.5 μm range and high thermal emissivity within the atmospheric transparency window (8–14 μm). However, to realize visible coloration, selective absorption within the visible spectrum (0.4–0.7 μm) is required. Incorporating vivid colors while maintaining efficient cooling is challenging, as it demands careful spectral engineering to minimize the trade-off between optical absorption in the visible range and cooling performance. The fundamental subtractive primary colors-cyan, magenta, and yellow (CMY)-are characterized by distinct reflection minima located approximately at 700 nm, 546.1 nm, and 435.8 nm, respectively [18]. In this work, the structure is designed to achieve high-contrast CMY colors through precise control of spectral reflectance while preserving the desired radiative cooling properties. Selective reflection dips generated by the optimized DBR and $MgF_2$ spacer produce the structural colors, whereas the PDMS layer and Ag back reflector maintain high thermal emissivity and strong solar reflectance, respectively. This integrated design minimizes the trade-off between color generation and cooling performance.

To quantitatively determine the perceived color of the proposed structure, the calculated reflection spectrum is converted into a corresponding color representation using the CIE colorimetric framework. The perceived color of a surface depends on the spectral power distribution of the illumination source, the wavelength-dependent reflectance of the material, and the spectral sensitivity of the human visual system. In this work, the standard illuminant D65 is used to represent natural daylight conditions [34]. The reflected spectral intensity is obtained by multiplying the incident light spectrum with the reflectance of the structure. This reflected spectrum is then mapped into the CIE 1931 XYZ color space using the standard color matching functions, which characterize the response of the human eye to different wavelengths.

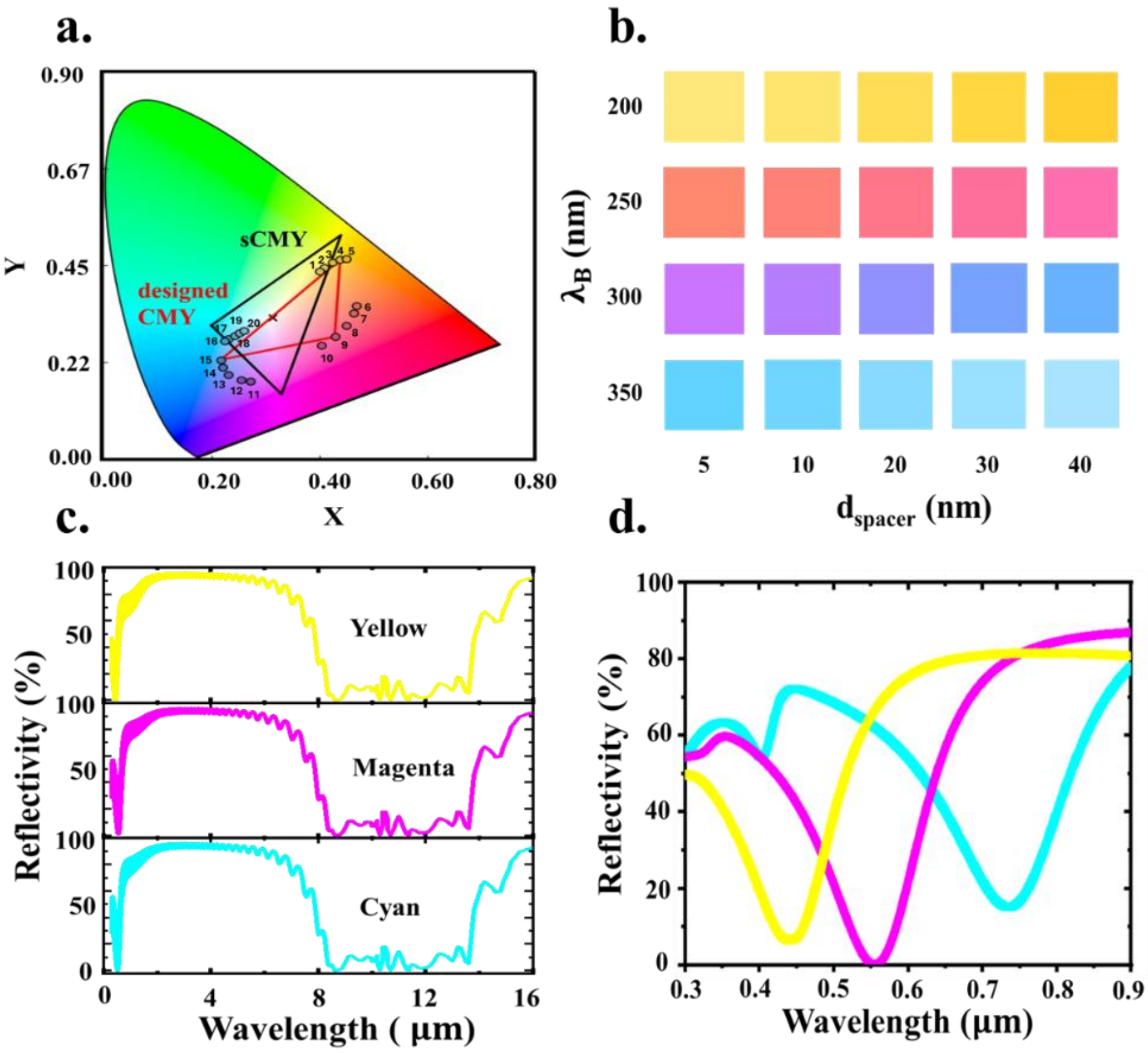

**Fig. 4.** (a) CIE 1931 chromaticity diagram comparing standard CMY and designed CMY colors, indicating the achievable color range. (b) Color tunability with variation of DBR Bragg wavelength (200–350 nm) and $MgF_2$ spacer thickness (5–40 nm). (c) Spectral reflectance of the designed CMY structures showing high NIR reflectivity and low reflectivity in the atmospheric window. (d) Enlarged visible region highlighting reflection dips at ~440 nm, 560 nm, and 740 nm corresponding to yellow, cyan, and magenta, respectively.

The tristimulus values $X$, $Y$, and $Z$ are calculated as [34]:

$$X = 100\frac{\int R(\lambda)\, I(\lambda)\, \bar{x}(\lambda)\, d\lambda}{\int I(\lambda)\, \bar{y}(\lambda)\, d\lambda}, Y = 100\frac{\int R(\lambda)\, I(\lambda)\, \bar{y}(\lambda)\, d\lambda}{\int I(\lambda)\, \bar{y}(\lambda)\, d\lambda}, Z = 100\frac{\int R(\lambda)\, I(\lambda)\, \bar{z}(\lambda)\, d\lambda}{\int I(\lambda)\, \bar{y}(\lambda)\, d\lambda} \quad (10)$$

where $R(\lambda)$ is the spectral reflectance of the structure, $I(\lambda)$ is the spectral power distribution of the incident light, and $\bar{x}(\lambda)$, $\bar{y}(\lambda)$, and $\bar{z}(\lambda)$ are the CIE color matching functions.

The chromaticity coordinates are obtained by normalizing the tristimulus values as:

$$x = \frac{X}{X+Y+Z}, y = \frac{Y}{X+Y+Z}, z = \frac{Z}{X+Y+Z} \quad (11)$$

where $x + y + z = 1$. Therefore, only the (x, y) coordinates are sufficient to represent the perceived color on the CIE 1931 chromaticity diagram. To evaluate perceptual color differences more effectively, the XYZ values are further transformed into the CIE L*a*b* color space, which provides a more uniform representation of human color perception. The conversion is given by:

$$
\begin{aligned}
L^* &= 116\, f\left(\frac{Y}{Y_n}\right) - 16 \\
a^* &= 500\left[f\left(\frac{X}{X_n}\right) - f\left(\frac{Y}{Y_n}\right)\right] \\
b^* &= 200\left[f\left(\frac{Y}{Y_n}\right) - f\left(\frac{Z}{Z_n}\right)\right]
\end{aligned}
\tag{12}
$$

where $X_n$, $Y_n$, and $Z_n$ are the tristimulus values of the reference white under D65 illumination. The function $f(t)$is defined as:

$$
f(t) = \begin{cases} t^{\frac{1}{3}}, & t > \left(\frac{6}{29}\right)^3 \\ \dfrac{t}{3\left(\frac{6}{29}\right)^2} + \dfrac{4}{29}, & \text{otherwise} \end{cases}
\tag{13}
$$

Finally, the color difference between two samples is evaluated using the Euclidean distance in the CIE L*a*b* space:

$$
\Delta E = \sqrt{(\Delta L^*)^2 + (\Delta a^*)^2 + (\Delta b^*)^2}
\tag{14}
$$

This metric provides a quantitative measure of perceptual color variation and is used to optimize the structural parameters for accurate color reproduction. Based on the calculated CMY color coordinates, the corresponding CIE 1931 chromaticity diagram is presented in Fig. 4(a), where the standard CMY (sCMY) and the colors obtained from the proposed structure (designed CMY) are compared. The vertices of the black triangle represent the ideal subtractive primary colors (CMY), while the red triangle corresponds to the colors achieved by the proposed PDCRC. The area enclosed by these triangles provides a measure of the achievable color gamut, indicating the color selectivity and contrast of the designed structure. Although the designed CMY points do not exactly coincide with the standard CMY coordinates, the structure still demonstrates a broad color range with high visual vividness.

The color generation is governed by the spectral response of the color-selective layers, as illustrated in Fig. 1(c). As shown in Fig. 4(b), the reflected color can be tuned from yellow to cyan by varying the Bragg wavelength of the distributed Bragg reflector (DBR) from 200 to 350 nm. The DBR layer thicknesses ($SiO_2$ and SiC) are related to the Bragg wavelength through the quarter-wave condition, given by $t = \lambda_{\mathrm{DBR}}/(4n)$, where $n$ is the refractive index of the respective material [18]. In addition, the color contrast can be further modulated by adjusting the thickness of the $MgF_2$ spacer layer in the range of 5–40 nm. These variations lead to shifts in the reflection dip within the visible spectrum, thereby enabling precise control over the perceived color.

The full spectral reflectance profiles of the three primary colors are shown in Fig. 4(c). The structure exhibits high reflectivity in the NIR region (0.75–6 μm) and low reflectivity (high emissivity) within the atmospheric window (8–14 μm), while maintaining distinct reflection dips in the visible range. These reflection minima are more clearly illustrated in Fig. 4(d), where dips are observed at approximately 440 nm, 560 nm, and 740 nm for yellow, cyan, and magenta, respectively. Although these wavelengths are slightly shifted compared to the ideal sCMY values, they still produce well-defined and distinguishable colors.

## 4.2. Optimization of DBR and Spacer Thickness Using Color Difference Analysis

To achieve high-purity CMY colors, the deviation of the designed colors from the standard CMY (sCMY) values is systematically analyzed. This is quantified using the color difference metric (ΔE) between the designed and standard colors. The optimal DBR(SiC/$SiO_2$) and spacer ($MgF_2$) thicknesses are determined by minimizing ΔE. The optimization is performed iteratively by varying one parameter while keeping the other fixed. Initially, the DBR Bragg wavelength ($\lambda_B$) is varied while maintaining a constant $MgF_2$ thickness. From Fig. 4(b), it is observed that a Bragg wavelength of 350 nm corresponds to cyan. Accordingly, ΔE is evaluated over a $\lambda_B$ range of 310–400 nm, as shown in Fig. 5(a). A minimum ΔE of 24 is obtained at $\lambda_B \approx 380$ nm, which is selected as the optimized value. The corresponding color variations are illustrated on the CIE 1931 diagram in Fig. 5(b). Although the color at 380 nm appears slightly faded, a range of $\lambda_B$ values between 330 and 380 nm can be considered depending on the desired color intensity. Next, with $\lambda_B$ fixed at 380 nm, the $MgF_2$ spacer thickness is varied. As shown in Fig. 5(c), two local minima in ΔE are observed near 15 nm and 55 nm, corresponding to relatively deeper and lighter cyan shades, respectively (Fig. 5(d)).

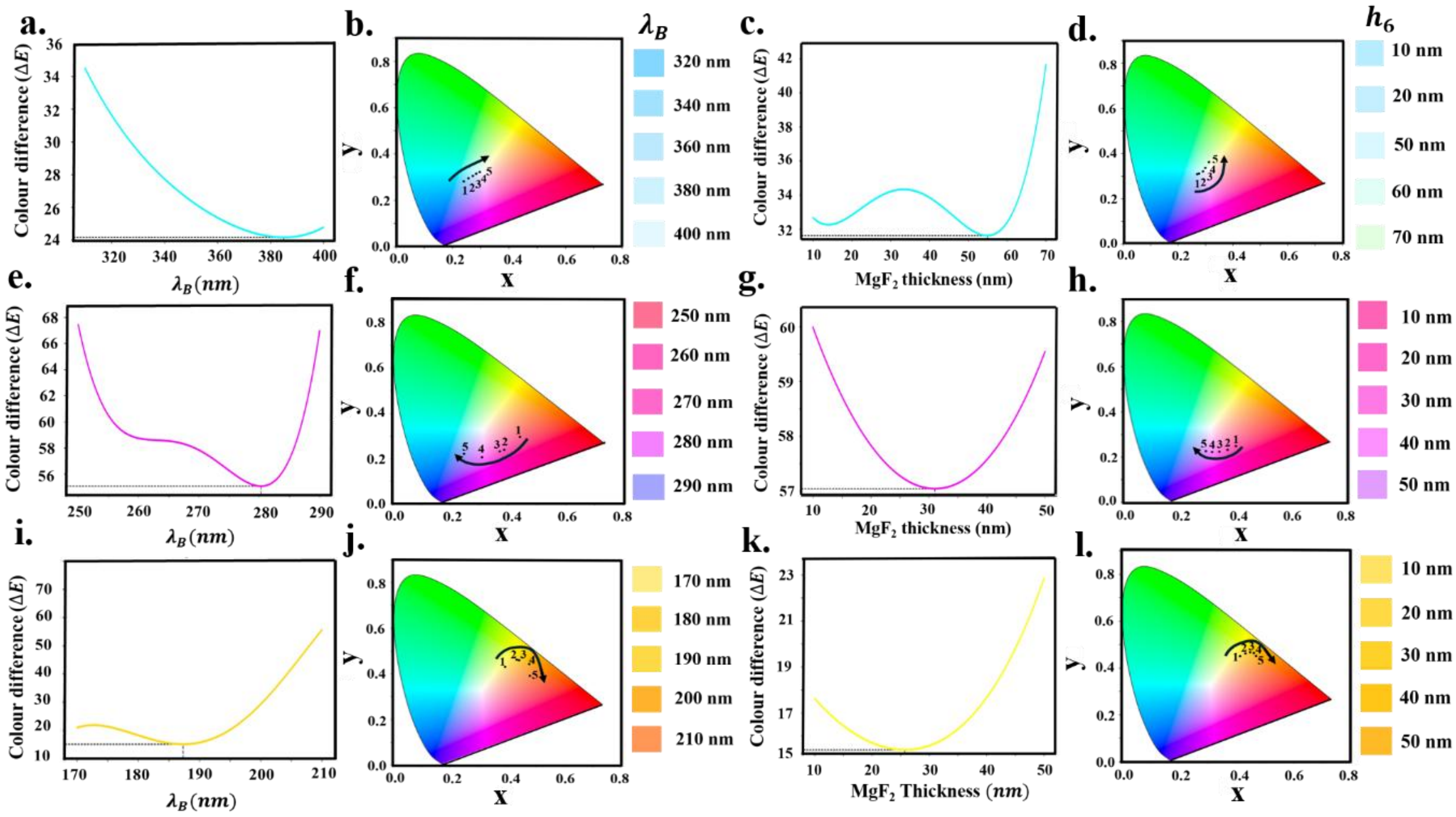

**Fig. 5.** Optimization of CMY colors based on color difference (ΔE) analysis. (a–d) Cyan optimization: ΔE variation with DBR Bragg wavelength in the range of 310–400 nm and $MgF_2$ spacer thickness of 5–60 nm, showing a minimum ΔE ≈ 24 at $\lambda_B \approx 380$ nm, along with corresponding color evolution. (e–h) Magenta optimization: ΔE variation for $\lambda_B$ ranging from 250–290 nm and spacer thickness of 10–70 nm, with a minimum ΔE ≈ 57 at $\lambda_B \approx 280$ nm and spacer thickness ≈ 42 nm. (i–l) Yellow optimization: ΔE variation for $\lambda_B$ between 170–210 nm and spacer thickness of 10–50 nm, showing a minimum ΔE ≈ 14 at $\lambda_B \approx 187$ nm and spacer thickness ≈ 25 nm.

To optimize for magenta, the Bragg wavelength is varied from 250 to 290 nm (Fig. 5(e)), and the corresponding color evolution is shown in Fig. 5(f). A pronounced minimum in ΔE is observed at $\lambda_B \approx 280$ nm, which is selected as the optimized value. Subsequently, by fixing this wavelength, the $MgF_2$ thickness is varied from 10 to 70 nm. As shown in Fig. 5(g), a minimum ΔE of approximately 57 is obtained at a spacer thickness of ~42 nm. The corresponding color variation (Fig. 5(h)) indicates a transition from deep to lighter magenta with increasing spacer thickness. For yellow, the Bragg wavelength is varied, and the corresponding ΔE profile is presented in Fig. 5(i). A minimum ΔE of 14 is achieved at $\lambda_B \approx 187$ nm, which is selected as the optimized value. The associated color variation (Fig. 5(j)) shows a transition from light to deep yellow within the explored range. Further optimization of the $MgF_2$ thickness (Fig. 5(k)) reveals a minimum ΔE at approximately 25 nm. Finally, based on the overall comparison of cyan, magenta, and yellow cases (Fig. 5(d), (h), and (l)), an $MgF_2$ spacer thickness of 20 nm is selected as a practical compromise to achieve improved color contrast and visual clarity across all three primary colors.

## 4.3. Angular and Polarization Dependence of Spectral Emissivity and Color

Since solar radiation impinges on the surface of the PDCRC over a wide range of incident angles and polarization states, it is essential to evaluate the optical performance of the structure under such conditions. Figure 6(a) illustrates the incident configuration, where the angle of incidence (θ) and polarization angle (φ) define the transverse electric (TE) and transverse magnetic (TM) polarization components. The emissivity profiles of the three colored radiative coolers at different incident angles are presented in Fig. 6(b). At normal incidence (0°), the PDCRC exhibits a high average emissivity of approximately 92%. As the incident angle increases, the emissivity gradually decreases to 87.5% at 10°, 78.2% at 20°, and 63% at 30°, eventually reaching 29.8% at 60°. This behaviour indicates that the PDMS-based emitter performs most efficiently under near-normal incidence, corresponding to peak solar conditions around midday, while its performance reduces at larger angles.

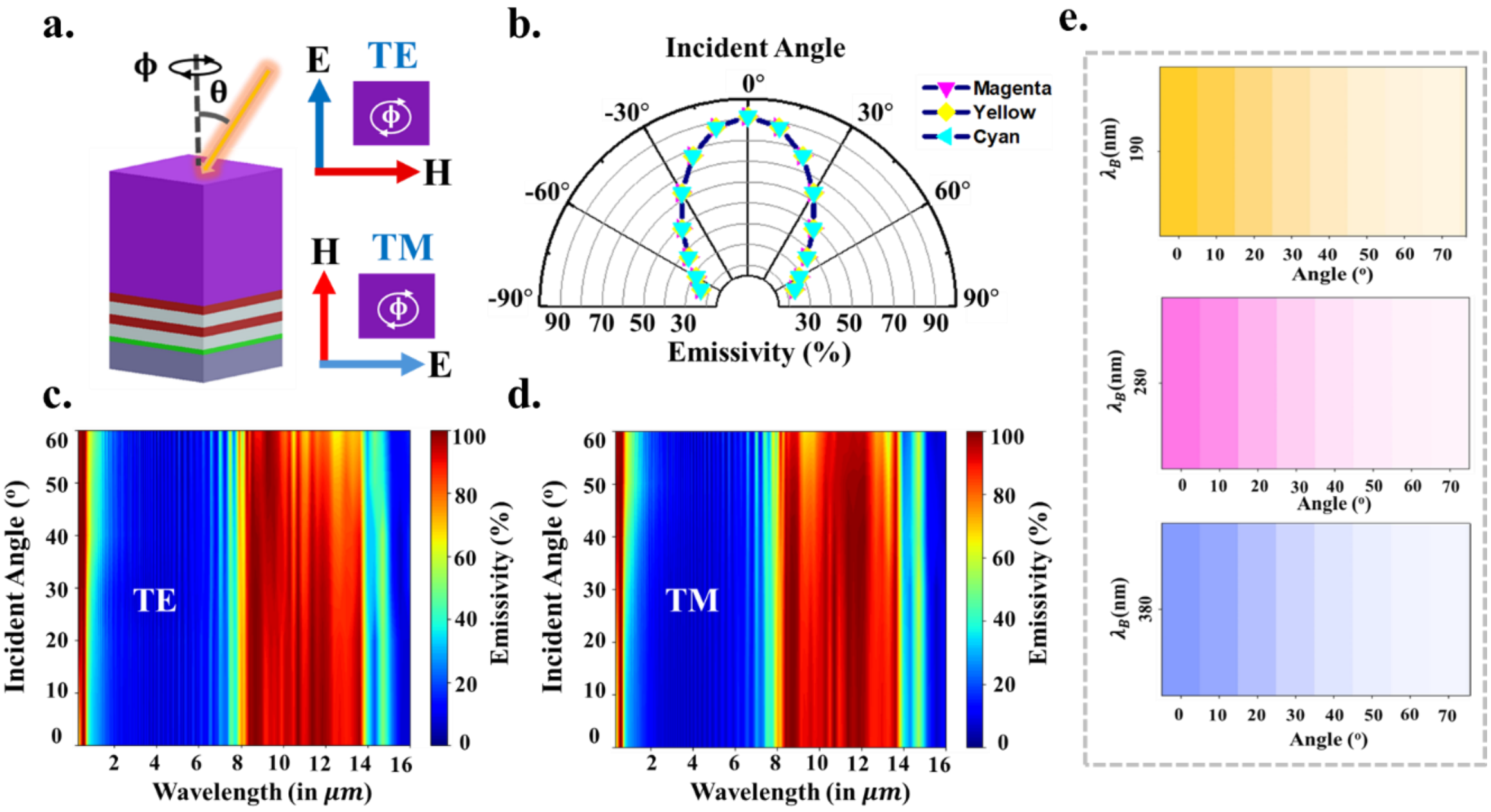

**Fig. 6.** (a) Schematic illustration of incident light showing the angle of incidence (θ) and polarization states (TE and TM). (b) Variation of average emissivity of the PDCRC with incident angle, decreasing from ~92% at normal incidence to ~78.2% at 20° and ~29.8% at 60°. (c,d) Spectral emissivity at different incident angles for TE and TM polarizations, respectively, showing nearly identical responses and indicating polarization-independent behaviour, particularly beyond 50° near 14 μm. (e) Angular dependence of the reflected CMY colors, exhibiting high contrast at normal incidence with gradual fading toward ~70°.

The spectral emissivity variations over the full wavelength range for TE and TM polarizations are shown in Fig. 6(c) and 6(d), respectively. The emissivity profiles for both polarizations exhibit nearly identical behaviour, with only minor deviations observed at higher incident angles. In particular, beyond 50° incidence, the emissivity near 14 μm shows negligible variation between TE and TM modes, indicating that the structure exhibits polarization-insensitive characteristics. Figure 6(e) presents the angular dependence of the perceived CMY colors. It is observed that the colors display high contrast and saturation at normal incidence, while gradually fading as the incident angle increases toward 70°. This reduction in color intensity arises from the angle-dependent shift and broadening of the reflection dip in the visible spectrum, which weakens the selective reflection responsible for structural coloration. Consequently, although the color purity decreases at larger incident angles, the dominant hue remains unchanged, demonstrating good angular color stability.

## 4.4. Cooling Performance of the Passive Daytime Colored Radiative Cooler

In order to estimate the cooling performance of the proposed structure, we calculate the net cooling power as a result of the total energy balance between the incoming and outgoing radiative and non-radiative heat fluxes. The net cooling power per unit area, $P_{\text{net}}(T_s)$, is given by [35]:

$$P_{\text{net}}(T_s) = P_{\text{rad}}(T_s) - P_{\text{atm}}(T_{\text{amb}}) - P_{\text{sun}} - P_{\text{non-rad}} \tag{15}$$

where $T_s$ and $T_{\text{amb}}$ are the sample temperature of the sample and the ambient temperature, respectively. The term $P_{\text{rad}}(T_s)$ represents the radiated power from the structure, while $P_{\text{atm}}$ and $P_{\text{sun}}$ correspond to the absorbed atmospheric and solar power, respectively. The term $P_{\text{non-rad}}$ represents for heat exchange due to conduction and convection.

The radiative power emitted is calculated by integrating the blackbody radiation multiplied by the spectral emissivity over wavelength and angle:

$$P_{\text{rad}}(T_s) = \int_0^\infty \int_\Omega \varepsilon(\lambda,\theta)\ I_{\text{bb}}(T_s,\lambda)\ \cos\theta\ d\Omega\ d\lambda \tag{16}$$

where $\varepsilon(\lambda,\theta)$ denotes the spectral directional emissivity of the structure, and $I_{\text{bb}}(T,\lambda)$ is the blackbody radiation spectra given by Planck's law :

$$I_{\text{bb}}(T,\lambda) = \frac{2hc^2}{\lambda^5}\frac{1}{\exp\left(\frac{hc}{\lambda k_B T}\right) - 1} \tag{17}$$

where, $h$, $c$, and $k_B$ represent Planck's constant, the speed of light in vacuum, and Boltzmann's constant, respectively. Similarly, the absorbed atmospheric radiation is calculated as:

$$P_{\text{atm}}(T_{\text{amb}}) = \int_0^\infty \int_\Omega \varepsilon(\lambda,\theta)\ \varepsilon_{\text{atm}}(\lambda,\theta)\ I_{\text{bb}}(T_{\text{amb}},\lambda)\ \cos\theta\ d\Omega\ d\lambda \tag{18}$$

where $\varepsilon_{\text{atm}}(\lambda,\theta)$ is the directional spectral emissivity of the atmosphere. It is related to the atmospheric transmittance $t(\lambda)$ through [36]:

$$\varepsilon_{\text{atm}}(\lambda,\theta) = 1 - t(\lambda)^{\frac{1}{\cos\theta}} \tag{19}$$

where $t(\lambda)$ is the atmospheric transmittance in the zenith direction. The atmospheric emissivity is low within the 8–14 µm wavelength range due to weak absorption by atmospheric gases, resulting in high transmittance. In contrast, the designed structure exhibits high emissivity in this wavelength range, enabling efficient thermal radiation and thereby facilitating effective passive radiative cooling.

The absorbed solar power is evaluated over the solar spectrum as:

$$P_{\text{sun}} = \int_{0.3\ \mu m}^{2.5\ \mu m} \varepsilon(\lambda)\ I_{\text{solar}}(\lambda)\ d\lambda \tag{20}$$

where $I_{\text{solar}}(\lambda)$ corresponds to the incident solar irradiance spectrum. Under normal incidence, the spectral emissivity is approximated as $\varepsilon(\lambda) = 1 - R(\lambda)$, where $R(\lambda)$is the reflectance of the structure.

The non-radiative heat transfer is modeled as:

$$P_{\text{non-rad}} = h_c(T_s - T_{\text{amb}}) \tag{21}$$

where $h_c$ is the effective heat transfer coefficient accounting for convection and conduction losses. Based on these formulations, the cooling performance of the device is quantified using two key metrics: (i) the equilibrium temperature drop, obtained when $P_{\text{net}} = 0$, and (ii) the net cooling power at ambient conditions when $T_s = T_{\text{amb}}$. These parameters provide a comprehensive assessment of the cooling capability of the proposed colored radiative cooler under realistic operating conditions.

The cooling performance of the proposed passive daytime colored radiative cooler (PDCRC) is illustrated in Fig. 7. Although high reflectivity in the visible-to-near-infrared (Vis–NIR) region helps suppress solar heating, the overall cooling capability is primarily governed by the thermal emissivity of the structure within the atmospheric transparency window. Since the DBR layers and Ag back reflector mainly contribute to spectral reflection rather than thermal emission, the emissivity dependence on PDMS and $MgF_2$ thickness is systematically investigated, as shown in Fig. 7(a). The emissivity of the structure, represented by the green markers, increases gradually with PDMS thickness. Average emissivity values of approximately 66.5%, 77%, 84.8%, 87.5%, 89.7%, and 92 % are obtained for PDMS thicknesses of 5, 10, 15, 20, 25, and 30 μm, respectively. Beyond 30 μm, the emissivity approaches saturation, reaching a maximum value of nearly 93−94% for PDMS thicknesses of 40–50 μm. Therefore, a PDMS thickness of 30 μm is selected as an optimal compromise between emissivity enhancement and structural practicality. In contrast, the emissivity represented by the orange markers remains nearly unchanged as the $MgF_2$ spacer thickness varies from 5 to 50 nm, indicating that the spacer layer has a negligible influence on the thermal emissivity of the structure.

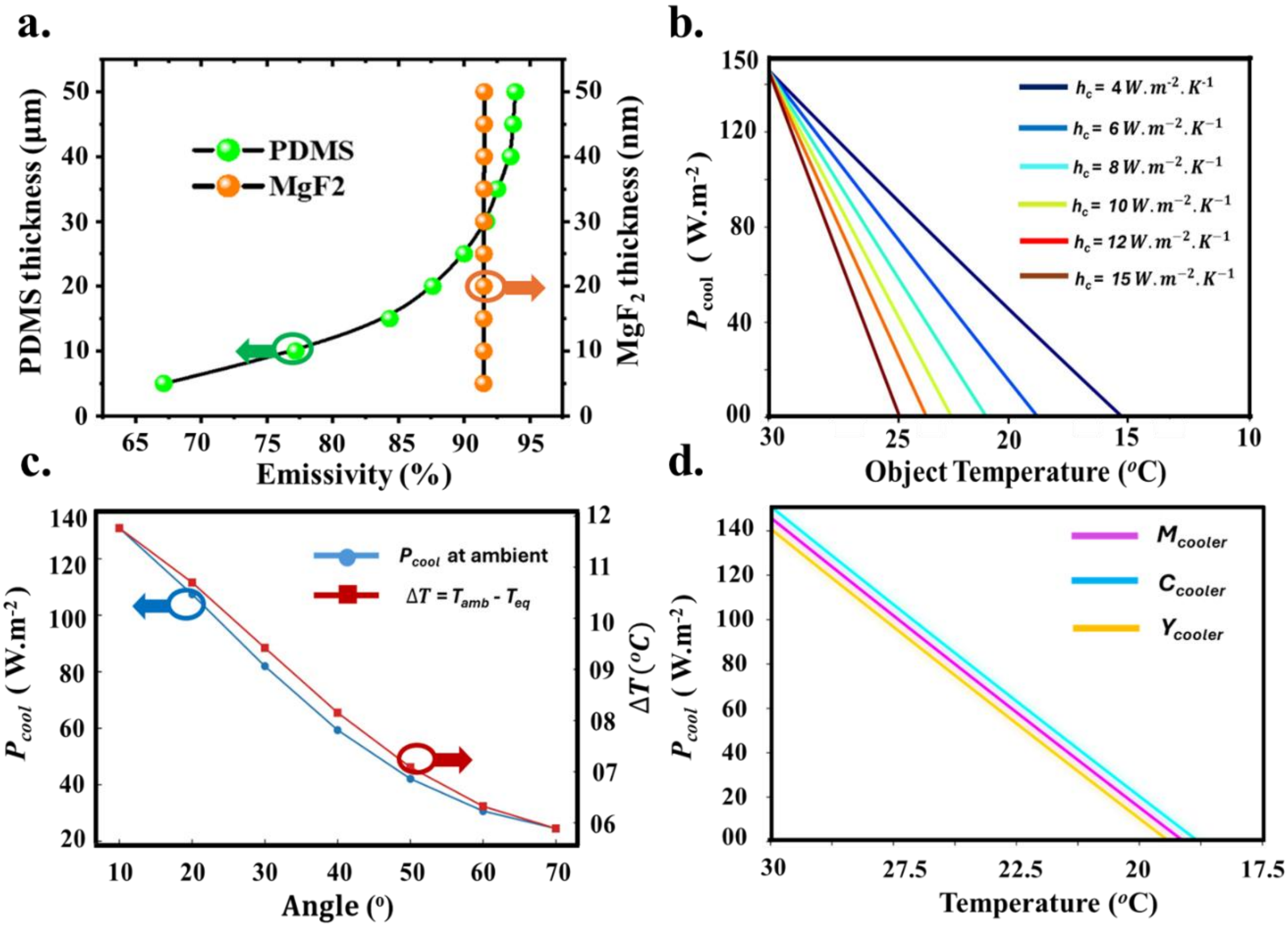

**Fig. 7.** (a) Variation of average emissivity with PDMS thickness (5–50 μm) and $MgF_2$ spacer thickness (5–50 nm), showing emissivity saturation beyond 30 μm PDMS thickness and negligible dependence on $MgF_2$ thickness. (b) Net cooling power of the PDCRC under different non-radiative heat transfer coefficients ($h_c = 4$–$15\ \mathrm{W\ m^{-2}K^{-1}}$) under normal solar incidence showing equilibrium cooling temperatures and maximum cooling power of ~145 $\mathrm{W\ m^{-2}}$. (c) Variation of cooling power and temperature reduction ($\Delta T$) with incident angle from 10° to 70°, for practical outdoor conditions ($h_c = 6\ \mathrm{W\ m^{-2}K^{-1}}$) indicating gradual degradation of cooling performance at larger incident angles. (c) Cooling performance comparison of cyan, magenta, and yellow radiative coolers, exhibiting cooling powers above 140 $\mathrm{W\ m^{-2}}$ ($h_c = 6\ \mathrm{W\ m^{-2}K^{-1}}$) and equilibrium temperatures around 18–19°C.

Figure 7(b) presents the calculated net cooling power of the PDCRC under different non-radiative heat transfer coefficients ($h_c$) ranging from 4 to 15 $\mathrm{W\ m^{-2}K^{-1}}$ under normal solar incidence. The vertical axis represents the net cooling power, while the horizontal axis denotes the equilibrium temperature of the cooler. The equilibrium cooling temperature corresponds to the condition where the net cooling power becomes zero ($P_{\mathrm{cool}} = 0$). The maximum cooling power achieved by the proposed PDCRC is found to be ~ 145 $\mathrm{W\ m^{-2}}$. At $h_c = 4\ \mathrm{W\ m^{-2}K^{-1}}$, the structure achieves a maximum temperature reduction of nearly 14.8°C relative to the ambient temperature of 30°C, corresponding to an equilibrium temperature of approximately 15.2°C. As the non-radiative heat transfer coefficient increases, the cooling performance gradually decreases, resulting in a temperature reduction of about 5°C at $h_c = 15\ \mathrm{W\ m^{-2}K^{-1}}$. For a practical outdoor condition of $h_c = 6\ \mathrm{W\ m^{-2}K^{-1}}$ [11], [18], the proposed structure demonstrates a cooling capability of approximately 12°C below ambient temperature.

Figure 7(c) illustrates the variation of cooling power and temperature reduction ($\Delta T$) from the ambient temperature ($T_{\text{amb}}$) to the equilibrium temperature ($T_{\text{eq}}$) as a function of incident angle ranging from 10° to 70°. It is observed that both the cooling power and temperature reduction gradually decrease with increasing angle of incidence. At normal incidence, the cooling power exceeds 140 $\text{W m}^{-2}$, as discussed previously. For an incident angle of 10°, the cooling power decreases to approximately 135 $\text{W m}^{-2}$ with a corresponding temperature reduction of $\Delta T \approx$ 11.7°C. As the incident angle increases to 20°, the cooling power further decreases to nearly 108 $\text{W m}^{-2}$, accompanied by a temperature reduction of about 10.8°C. For incident angles of 30° and 40°, the cooling powers are approximately 83 $\text{W m}^{-2}$ and 59 $\text{W m}^{-2}$, respectively, while the corresponding temperature reductions decrease to 9.6°C and 8.2°C. At a large incident angle of 70°, the cooling power reduces significantly to nearly 24 $\text{W m}^{-2}$ with a temperature reduction of approximately 5.8°C. This reduction in cooling performance at higher incident angles can be attributed to the simultaneous decrease in both effective emissivity and incident solar intensity [37], [38]. The cooling performance of the cyan, magenta, and yellow radiative coolers is compared in Fig. 7(d). All three structures exhibit cooling powers exceeding 140 $\text{W m}^{-2}$, with cyan showing the highest cooling power of approximately 145 $\text{W m}^{-2}$, followed by magenta (~143 $\text{W m}^{-2}$) and yellow (~140 $\text{W m}^{-2}$). The equilibrium temperatures of all three colored radiative coolers remain within the range of 18–19°C, corresponding to a temperature reduction of approximately 11–12°C below the ambient condition. These results confirm that the proposed structural coloration strategy preserves strong radiative cooling performance while simultaneously enabling visually distinguishable CMY colors.

**Table1:** Comparative performance of reported radiative cooling structures and the proposed PDMS/DBR/MgF2/Ag design

| **Ref.** | **Structure** | **Color Generation Mechanism** | **Solar Reflectivity (%)** | **Average Emissivity (8–14 µm) (%)** | **Cooling Power ($\text{W m}^{-2}$)** | **ΔT (°C)** | **Fabrication Method** |
|---|---|---|---|---|---|---|---|
| [11] | $HfO_2/SiO_2$/Ag/Ti/Si | No color | 97 | | 40.1 | 5 | Multilayer deposition |
| [16] | Quartz array/Si/Al | Periodic photonic structure | – | – | – | 14.4 | Lithography |
| [17] | $SiO_2/Si_3N_4$/Ag/Si | Optical resonance | – | – | – | 3.9 | Multilayer deposition |
| [18] | Tamm structure | Optical Tamm resonance | – | 80 | 44–52 | 5–6 | Multilayer deposition |
| [19] | PDMS-Polystyrene microsphere | Particle scattering | – | 96 | – | 4 | Solution processing |
| [20] | $SiO_2/TiO_2$ periodic structure | Plasmonic resonance | 98 | – | ~60 | – | Nanopatterning |
| [21] | Hole-array photonic structure | Photonic crystal resonance | – | 98.39 | 139 | – | Nanofabrication |
| [34] | $SiO_2$ meta/$TiO_2$/PDMS/Ag | Polarization controlled | – | 96.62 | 135.21 | – | Deposition and Metastructure |
| [36] | $SiO_2/TiO_2$/Ag | MDM structural based | – | – | 72.14 | 9.44 | Nine-layer deposition |
| **This work** | **PDMS/DBR/$MgF_2$/Ag** | **DBR-assisted structural color selectivity** | **93.3** | **92** | **140–145** | **~12** | **multilayer deposition** |

Table 1 provides a comparative overview of reported radiative cooling and photonic structures along with their corresponding color mechanisms, fabrication approaches, and cooling performance metrics. Earlier works based on the Tamm structure, SiO2/Si3N4-based thin-film resonators, PDMS-Polystyrene microsphere-based structures, and other plasmonic and multilayer structures demonstrated moderate sub-ambient cooling temperature ranging from 3.9°C to 9.44°C except for the quartz crystal based structure which achieved 14.4°C sub-ambient cooling theoretically, and cooling power ranging from 40.1-139 W $m^{-2}$. They are primarily governed by photonic interference, optical resonance, or scattering mechanisms. In comparison, the proposed PDMS/DBR/$MgF_2$/Ag multilayer structure exhibits superior solar reflectivity of 93.3% and high average emissivity of 92% within the atmospheric window, resulting in an enhanced cooling power of 140–145 W $m^{-2}$ and a temperature reduction of approximately 12°C. This superior performance originates from the synergistic spectral engineering of the multilayer structure, which simultaneously maximizes solar reflection, enhances thermal emission, and enables selective visible-light absorption for structural color generation. The optimized combination of the highly emissive PDMS layer, the SiC/$SiO_2$ distributed Bragg reflector for efficient solar reflection and color selectivity, the $MgF_2$ spacer for precise tuning of the reflection dip, and the Ag back reflector for suppressing optical transmission minimizes solar heat gain while maximizing radiative heat dissipation through the atmospheric transparency window. Consequently, the proposed design effectively minimizes the trade-off between color generation and daytime radiative cooling performance, highlighting its potential for scalable and cost-effective radiative cooling applications.

## 4.5. Tolerance analysis and Fabrication feasibility

To assess the fabrication robustness of the proposed lithography-free PDCRC, a thickness tolerance analysis was performed by independently varying the thickness of each constituent layer while keeping the remaining layers at their optimized values. The average emissivity was evaluated over the atmospheric transparency window (8–14 μm), whereas the average reflectivity was calculated over the wavelength range of 0.75–6 μm. The visible region (0.4–0.7 μm) was excluded from the reflectivity calculation because the intentionally introduced reflection dips responsible for CMY structural color generation reduce the reflectivity within this spectral range and do not contribute to parasitic solar absorption. As summarized in Table 2, the proposed structure exhibits excellent tolerance against practical fabrication deviations. Among all the constituent layers, the PDMS emitter has the most pronounced influence on the radiative cooling performance because it primarily governs thermal emission within the atmospheric transparency window. Increasing the PDMS thickness enhances the average emissivity, whereas the average reflectivity remains almost unchanged, indicating that the emitter thickness mainly affects infrared emission without influencing the solar-reflective characteristics.

In contrast, thickness variations in the $SiO_2$ and SiC layers introduce only minor changes in the average emissivity and reflectivity. These dielectric layers primarily determine the optical interference required for structural coloration and solar reflection; therefore, moderate fabrication deviations have little impact on the overall cooling performance. Similarly, the $MgF_2$ spacer exhibits negligible sensitivity to thickness variations, confirming that its principal role is to fine-tune the visible reflection spectrum for color optimization while preserving the radiative cooling characteristics. The Ag back reflector also maintains nearly identical emissivity and reflectivity over the investigated thickness range, demonstrating that the selected 200 nm thickness is sufficient to effectively suppress optical transmission.

The proposed PDCRC is compatible with conventional thin-film fabrication techniques and does not require any lithography or nanopatterning processes. The SiC, $SiO_2$, $MgF_2$, and Ag layers can be deposited using standard physical vapor deposition techniques such as RF magnetron sputtering or thermal evaporation, while the PDMS (10:1 base-to-curing-agent ratio) emissive layer can be readily prepared by spin coating followed by thermal curing. Furthermore, the thickness tolerance analysis demonstrates that moderate fabrication deviations have only a negligible influence on the optical performance, confirming the practicality and manufacturability of the proposed multilayer architecture. These features make the design well suited for large-area fabrication for practical daytime radiative cooling applications.

**Table2:** Thickness tolerance analysis of the proposed lithography-free PDCRC.

| Layer | Thickness (nm or µm) | Average Emissivity (8–14 µm) (%) | Average Reflectivity (0.75–6 µm) (%) |
|---|---|---|---|
| PDMS | 20 µm (-10 µm) | 87.5 | 93.2 |
| | 30 µm | 92 | 93.3 |
| | 40 µm (+10 µm) | 93 | 93.4 |
| $SiO_2$ | 50 nm (-10 nm) | 91.5 | 92.8 |
| | 60 nm | 92 | 93.3 |
| | 70 nm (+10 nm) | 91.8 | 93.5 |
| SiC | 20 nm (-5 nm) | 91.9 | 91.6 |
| | 25 nm | 92 | 93.3 |
| | 30 nm (+5 nm) | 91.7 | 92.7 |
| $MgF_2$ | 15 nm (-5 nm) | 92 | 93.2 |
| | 20 nm | 92 | 93.3 |
| | 25 nm (+5 nm) | 92 | 93.3 |
| Ag | 170 nm (-30 nm) | 92 | 93.3 |
| | 200 nm | 92 | 93.3 |
| | 230 nm (+30 nm) | 92 | 93.3 |

## 4.6. Comparative Cooling Performance across Different Global Locations

To demonstrate the practical applicability of the radiative cooler, its cooling performance is theoretically investigated for different global cities. For the cooling power calculations, the average monthly meteorological data for July 2025 are considered. The solar intensity spectrum and atmospheric transmittance data were obtained from the "NASA POWER Data Access" Database [39]. In the calculations, we consider $h_c = 6\ \mathrm{W\,m^{-2}\,K^{-1}}$, an ambient temperature of 30°C, and the average solar intensity and atmospheric transmittance for the selected month. The emissivity of the proposed radiative cooler is used to evaluate the net cooling power under realistic environmental conditions. Figure 8(a) presents a global map showing major cities distributed across different climatic regions to demonstrate the performance of the radiative cooler under diverse environmental conditions. Figure 8(b) illustrates the calculated cooling power at different global locations. Among the considered cities, New Delhi exhibits the highest cooling power of 107.9 $\mathrm{W\,m^{-2}}$, whereas San Francisco shows the lowest value of 40.2 $\mathrm{W\,m^{-2}}$. The variation in cooling performance among different cities is primarily attributed to differences in the local solar irradiance and atmospheric transmittance used in the calculations. Since the atmospheric transmittance data inherently reflect the local atmospheric conditions, including the effects of water vapor absorption, the calculated cooling performance naturally varies from one location to another. However, humidity was not treated as an independent input parameter in the present analysis. The cooling powers of Beijing, Melbourne, and Seoul are calculated to be 87.2, 97.2, and 82.2 $\mathrm{W\,m^{-2}}$, respectively, while the remaining cities exhibit comparatively lower cooling performance due to regional climatic variations.

The obtained results indicate that the proposed radiative cooler maintains effective cooling capability over a broad range of environmental conditions and geographical locations. Such stable cooling performance highlights the suitability of the designed structure for real-world daytime passive cooling applications, including building thermal management, outdoor cooling systems, and energy-efficient temperature regulation technologies.

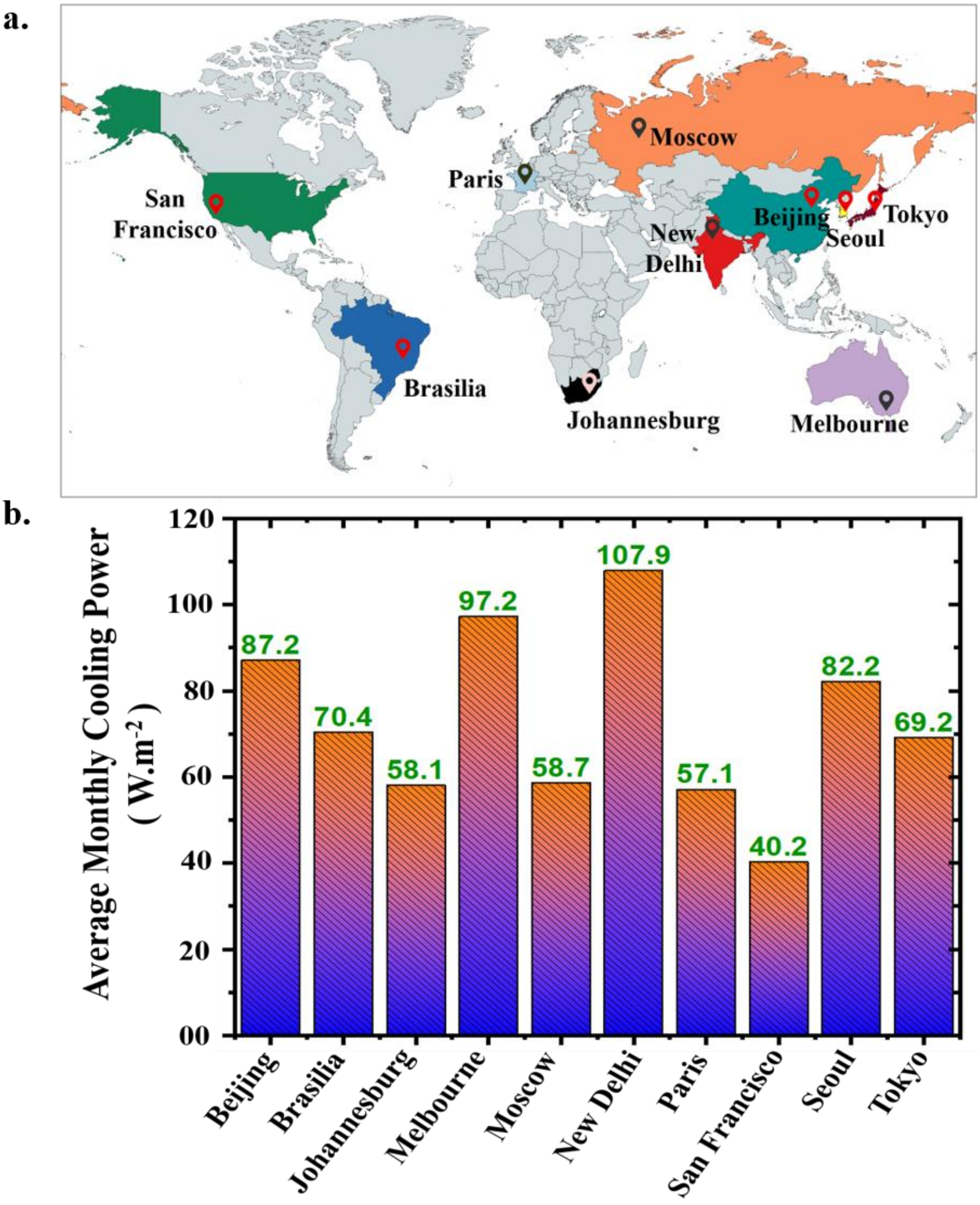

**Fig. 8.** (a) Global map showing the selected cities from different climatic regions used to evaluate the cooling performance of the proposed radiative cooler under realistic environmental conditions. (b) Calculated net cooling power of the PDCRC for different global cities using average meteorological data for July 2025, considering $h_c = 6\ \mathrm{W\ m^{-2}\ K^{-1}}$, ambient temperature of 30°C, and corresponding solar irradiance and atmospheric transmittance data. The highest cooling power of ~107.9 $\mathrm{W\ m^{-2}}$ is obtained for New Delhi, while the lowest value of ~40.2 $\mathrm{W\ m^{-2}}$ is observed for San Francisco.

## 5. Conclusion:

In summary, we proposed a multilayer lithography-free design of a passive daytime colored radiative cooler based on PDMS, $SiC/SiO_2$ DBR layers, an $MgF_2$ spacer, and an Ag back reflector and investigated it systematically. The structure simultaneously achieves selective visible-light absorption for color generation, high solar reflectance, and strong thermal emission within the atmospheric transparency window for cooling. By optimizing the DBR Bragg wavelength and spacer thickness using colour-difference analysis, high-contrast cyan, magenta, and yellow colors are successfully realized while maintaining excellent radiative cooling performance. The designed structure exhibits an average emissivity of approximately 92% within the 8-14 μm atmospheric window and a reflectivity of 93.3% in the near-infrared region. Electric-field distributions obtained from FDTD simulations and verified using the transfer matrix method reveal the transition from reflection-dominated behaviour at shorter wavelengths to strong thermal emission within the atmospheric window. Under practical operating conditions ($h_c = 6\,\mathrm{W\,m^{-2}\,K^{-1}}$), the proposed radiative cooler achieves cooling powers of 140–145 W $m^{-2}$ and a temperature reduction of approximately 12°C below ambient temperature. The structure also maintains stable cooling and color characteristics over a broad range of incident and polarization angles. In addition, cooling-power calculations using real climatic data from different global cities demonstrate the applicability of the proposed design under diverse environmental conditions. Compared with many previously reported colored radiative coolers that rely on complex nanostructures, metasurfaces, or lithography-based fabrication, the proposed design employs only planar thin-film layers and a single thick polymer emitter, making it more compatible with large-area and cost-effective fabrication techniques such as spin coating, sputtering, thermal evaporation, and other scalable thin-film deposition methods. This simplified architecture reduces fabrication complexity while preserving excellent optical and cooling performance. The combination of efficient passive cooling, vivid structural coloration, angular robustness, and fabrication simplicity makes the proposed PDCRC a promising candidate for practical thermal-management applications, including building facades in windows, vehicle coatings, wearable devices, and outdoor electronic systems. Future studies may focus on experimental realization, long-term environmental stability, self-cleaning surface integration, and adaptive color-tunable radiative cooling platforms to further enhance the practical deployment of colored radiative cooling technologies.

## Conflict of Interest

The authors declare that they have no conflict of interest.

## Ethics Statement

This research does not involve human participants, human or animal subjects, or any personally identifiable data. Therefore, ethical approval is not applicable.

## Credit of Authorship

**Rajib Lochan Ghadei:** Writing – Original Draft, Data Curation, Formal Analysis, Methodology, Writing – Review & Editing. **Rohit Gupta:** Data Curation, Formal Analysis, Methodology. **Rishi Maiti:** Funding Acquisition, Supervision, Writing – Review & Editing.

## Funding Statement

This work was supported by the Indian Institute of Technology Guwahati, Assam 781039, India.

## Data Availability Statement

The data supporting the findings of this study are available from the corresponding author upon reasonable request.